\documentclass[aps,prl,twocolumn,floatfix]{revtex4}
\usepackage{bm}
\usepackage{epsf}
\usepackage{amssymb}
\usepackage{amsmath}
\usepackage{mathbbol}
\usepackage{amsmath}
\usepackage{graphicx}
\usepackage{rotating}
\usepackage{epsfig}
\usepackage{psfrag}
\usepackage{amsmath}
\usepackage{hyperref}
\usepackage{setspace}
\usepackage{subfigure}

\newcommand{\Si}{{\rm Si}}

\DeclareMathOperator{\Tr}{Tr}

\begin{document}

\title{Why rare-earth ferromagnets are so rare: insights from the p-wave Kondo model}

\author{Shadab Ahamed$^{1,2}$, Roderich Moessner$^1$, Onur Erten$^1$}
\affiliation{$^1$Max-Planck-Institute for the Physics of Complex Systems, 01187 Dresden, Germany \\ $^2$Department of Physics, Indian Institute of Science, Bangalore 560 012, India}

\begin{abstract}
Magnetic exchange in Kondo lattice systems is of the Ruderman-Kittel-Kasuya-Yosida type, whose sign depends on the Fermi wave vector, $k_F$. In the simplest setting, for small $k_F$, the interaction is predominately ferromagnetic, whereas it turns more antiferromagnetic with growing $k_F$.
It is remarkable that even though $k_F$ varies vastly among the rare-earth systems, an overwhelming majority of  lanthanide magnets are in fact antiferromagnets. To address this puzzle, we investigate the effects of a p-wave form factor for the Kondo coupling pertinent to nearly all 
rare-earth intermetallics. We show that this leads to interference effects which for small $k_F$ are destructive, greatly reducing the size
of the RKKY interaction in the cases where ferromagnetism would otherwise be strongest.  By contrast, for large $k_F$, constructive interference can enhance antiferromagnetic exchange. Based on this, we propose a new route for designing ferromagnetic rare-earth magnets.
\end{abstract}

\maketitle

\noindent
{\it Introduction - } Magnetic exchange processes in quantum materials are intimately tied to their underlying electronic properties. For instance, most magnetic insulators in nature are antiferromagnets due to Anderson superexchange\cite{Anderson_PRB1950, Kanamori_ProgTheoPhys1957, Goodenough_JPhysChemS1958} whereas ferromagnets tend to be metallic and are stabilized by a combination of Hund's coupling and multiorbital effects\cite{Vollhardt1999, KatsnelsonRMP2008}. However, $f$-electron systems\footnote{By f-electron or rare-earth systems, we exclusively refer to Kondo lattice systems where the Kondo interaction is the dominant coupling between the local moment and the conduction electrons. We do not consider systems like elemental Gd, where the exchange is due to Hund's coupling.} provide a remarkably consistent exception to this rule: {\it a large majority of f-electron magnets are metallic antiferromagnets.} 
This is noteworthy since the Ruderman-Kittel-Kasuya-Yosida (RKKY) interaction\cite{Ruderman_PRB1954, Kasuya_PTP1956}, the dominant exchange mechanism, can be ferromagnetic or antiferromagnetic depending on the Fermi wave vector, $k_F$. Resolving this conundrum is particularly desirable in the context of the search  for Kondo lattice ferromagnets whose quantum criticality can lead to exotic superconductivity\cite{Huy_PRL2007, Aoki_Nature2001}.

The RKKY interaction results from a second order process where a local moment, $S_1$, first polarizes the conduction electrons through their Kondo coupling, $J_K$; this polarisation oscillates in space at a wavelength set by $k_F$, and its value at the location of another spin, $S_2$, in turn yields the magnetic exchange via its Kondo coupling. Historically, the Kondo interaction is considered to be onsite (or s-wave), with a local moment antiferromagnetically coupled to the local conduction electrons, Fig. 1(a). However, such an on-site Kondo coupling is absent in materials if  local moment and conduction electron wavefunctions have different symmetries. Therefore the interaction has to couple local and conduction electrons on neighboring sites. This is not a new insight\cite{Ikeda_JPSJ1996, Coleman_PRB1999}; however its importance has only been appreciated recently with the discovery of topological Kondo insulators\cite{Dzero_PRL2010, Alexandrov_PRL2013, Alexandrov_PRB2014, Alexandrov_PRL2015}, where the form factor plays a crucial role for the topological properties. The form of the interaction then depends on the angular momentum difference, $| \Delta l |$, between the local moment and the conduction electrons. 

We start from the observation that the the majority of Kondo lattice systems have conduction electrons derived from $d$ orbitals, yielding a $|\Delta l|=1$, p-wave, form factor, 

\begin{eqnarray}
\Phi({\bf \hat{r}}_{ij})^{\alpha\beta} = \frac{-i}{2}\sum_{\langle ij \rangle} {\boldsymbol \sigma}_{\alpha\beta} \cdot {\bf \hat{r}}_{ij} \ .
\label{formfactor}
\end{eqnarray}
Here ${\bf \hat{r}}_{ij} = ({\bf \hat{r}}_i-{\bf \hat{r}}_j$), coupling to the Pauli matrices is due to strong spin-orbit coupling in f orbitals. Eq. \ref{formfactor} can be derived from a periodic Anderson model with $s$ and spin-orbit coupled $p$ (j=1/2) orbitals. In a hypercubic lattice, the p-wave form factor couples nearest neighbors with different signs as shown in Fig. 1(c).  Note that even though $\Phi({\bf R})$ can take other forms, we mainly focus on the p-wave form factor which is pertinent to f-electron systems [see supplemental material for the extended s-wave form factor, Fig. 1.(b)].
\begin{figure}[t!]
\vspace{0.1cm}
\centerline{
\includegraphics[width=4.5cm]{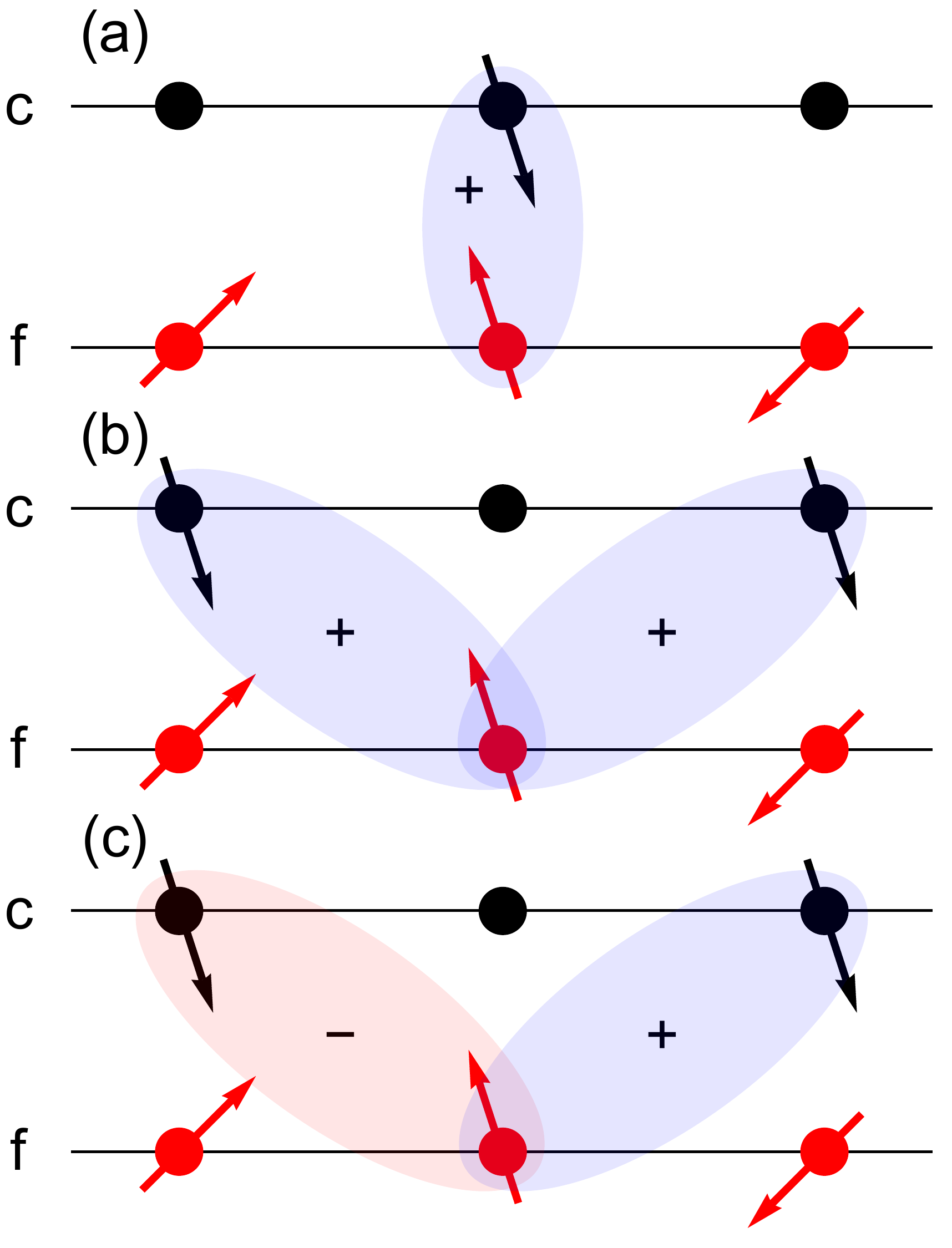}
}
\caption{
1D schematics of form factors for the Kondo interaction: (a) s-wave leads to the standard onsite interaction, (b) extended s-wave couples nearest neighbors with the same sign, (c) p-wave couples nearest neighbors with opposite signs.}
\label{fig1}
\end{figure}

\begin{figure*}[t!]
\vspace{0.1cm}
\centerline{
\includegraphics[width=17.5cm]{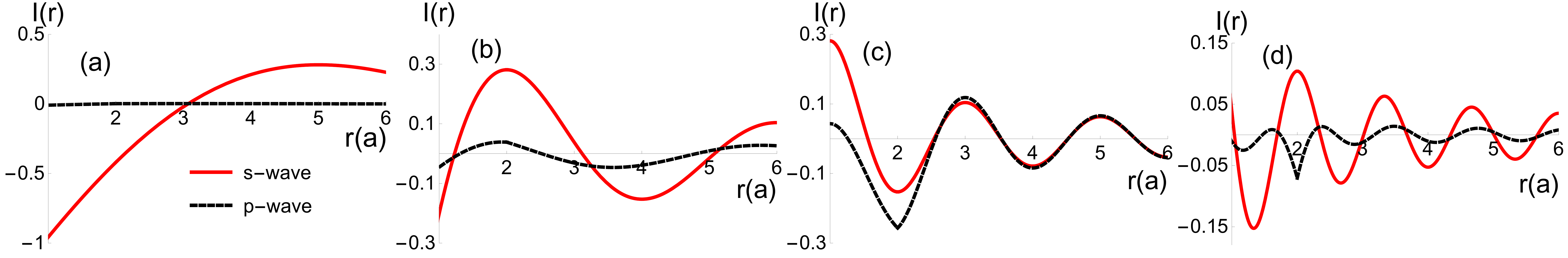}
}
\caption{
1D RKKY interactions $\mathcal{I}(r)$ in units of $J_K^2(k_Fa)^2/8\pi \epsilon_F$ for s-wave (red, solid), p-wave (black, dashed). (a) $k_Fa=\pi/10$, (b) $k_Fa=\pi/4$,(c) $k_Fa=\pi/2$ and (d) $k_Fa=3\pi/4$. For $k_Fa=\pi/10$, the interaction effectively vanishes for the p-wave case due to destructive interference. However for $k_Fa=\pi/2$, constructive interference stabilizes the exchange, particularly for $r>2a$.}
\label{fig2}
\end{figure*}
In this work, we investigate the role of the p-wave Kondo coupling for the RKKY interaction to address the puzzle, why the majority of Kondo lattice magnets are antiferromagnets. Our main results are: (i) RKKY interactions retain their Heisenberg form and spin-orbit coupling in Eq. \ref{formfactor} does not lead to anisotropic compass or Dzyaloshinskii-Moriya terms. (ii) Unlike the onsite Kondo coupling, the  p-wave case exhibits rich interference effects as a function of $k_F$. In general, the RKKY interaction is reduced, most pronouncedly so for small $k_F$ where ferromagnetism would have been strongest. (iii) Based on these insights, we propose a new route for the search of new rare-earth ferromagnets utilizing local inversion symmetry breaking to produce an extended s-wave form factor. We note that 
this local moment analysis does not directly apply to itinerant Kondo lattice systems such as ferromagnets like UCoGe\cite{Huy_PRL2007} and Sm$_2$Fe$_{12}$P$_7$\cite{Janoschek_JPCM2011}.

\noindent
{\it RKKY Model - } We start with the Kondo model
\begin{eqnarray}
H= \sum_{k,\alpha} \epsilon_k c^\dagger_{k\alpha}c_{k\alpha}+J_K\sum_i {\bf S}_i \cdot \psi_{i\alpha}^\dagger \sigma_{\alpha \beta} \psi_{i\beta}
\end{eqnarray} 
where $c^\dagger, c$ are the conduction electron fermion operators, ${\bf S}$ is the spin of the local moment. $\psi$ can be represented in terms of conduction electrons,
\begin{eqnarray}
\psi_{i\alpha}^\dagger = \Phi({\bf \hat{r}}_{ij})^{\alpha\beta} c^\dagger_{j\beta} 
\label{phi}
\end{eqnarray}
where $\Phi({\bf \hat{r}}_{ij})$ is the form factor, Eq. \ref{formfactor}. 
%For the s$_{\rm ext}$ wave form factor, we use $\Phi({\bf r}_{ij}) =(1/\sqrt{z})\sum_{\langle ij \rangle } \mathbb{1}$ which sums all nearest neighbors without a sign change or spin flip.
A canonical transformation\cite{Fazekas_1999} $\tilde{H}=e^{iS}He^{-iS}$ eliminates the $J_K$ term, generating an effective local moment exchange (see supplementary information for details). Collecting terms up to $J_K^2/\epsilon_F$, we obtain the effective RKKY interaction
\begin{eqnarray}
H_{RKKY} = \sum_{i\neq j} \mathcal{I}(r) {\bf S}_i \cdot {\bf S}_j 
\end{eqnarray}
with $r$ the distance between two sites, $r = |{\bf r}_i-{\bf r}_j| $ and $\mathcal{I}(r)$ a Lindhard function modified with the form factors
\begin{eqnarray}
 \mathcal{I}(r) = \frac{J_K^2}{8N^2}\sum_{{\bf kq}}\frac{e^{i({\bf k-q})\cdot \bf{r}_{ij}}}{\epsilon_k-\epsilon_q}  (F_k-F_q) \Tr({\Phi_k^\dagger \Phi_k}) \Tr({\Phi_q^\dagger \Phi_q}) 
 \label{Lindhard}
\end{eqnarray}
where $F_k$ is the Fermi-Dirac distribution function, $N$  the number of sites and $\Phi_k$ the Fourier transform of Eq. \ref{phi} such that $\psi_{k\alpha} = \Phi_k^{\alpha\beta}c_{k\beta}$ . Eq. \ref{Lindhard} holds generically, with $\Phi_k^\dagger = \Phi_k$
for hypercubic lattices. For the p-wave case, %$\Tr({\Phi_k^\dagger \Phi_k})$ is given as

\begin{eqnarray}
\Tr({\Phi_k^\dagger \Phi_k}) &=& 2\sum_{i=1}^D\sin(k_ia)^2
\end{eqnarray}
%\Tr({\Phi_k^\dagger \Phi_k})_{\rm s_{ext}} &=& [\sum_{i=1}^d\cos(k_i)]^2 
Ferromagnetic $\mathcal{I}(r)<0$  favors a $q=0$ state with all spins aligned, whereas $\mathcal{I}(r)>0$ favors antiparallel spin order. Next we evaluate eq. \ref{Lindhard} for 1D and 3D hypercubic lattices. In order to perform the integrals analytically, we use $\epsilon_k \simeq \hbar^2 k^2/2m$ and $\sum_{i=1}^D \sin(k_i)^2 \simeq \sin(k)^2$.

\noindent
{\it RKKY interaction in 1D -} Eq. \ref{Lindhard} in 1D yields for the onsite, s-wave case\cite{Yafet_PRB1987, Litvinov_PRB1998}
\begin{eqnarray}
\mathcal{I}_s^1(r)=\frac{J_K^2}{\epsilon_F}\frac{(k_Fa)^2}{8\pi} [\Si(2k_F r)-\pi/2]
\end{eqnarray}
with lattice constant $a$, sine integral function $\Si(x)$  and Fermi energy $\epsilon_F$. Similarly for the p-wave case, 

\begin{equation}\label{}
\mathcal{I}_p^1(r)=  \left\{\begin{array}{cc}
\frac{1}{16}[6\mathcal{I}_s^1(r)-4\mathcal{I}_s^1(r-a)-4\mathcal{I}_s^1(r+a)\\
+\mathcal{I}_s^1(r+2a)+\mathcal{I}_s^1(r-2a)], & r\geq 2a\\[0.15cm] 
 \frac{1}{16}[4\mathcal{I}_s^1(r)-4\mathcal{I}_s^1(r+a)+\mathcal{I}_s^1(r+2a)\\+\mathcal{I}_s^1(2a-r)-4\mathcal{I}_s^1(a)+2\mathcal{I}_s^1(2a)],& r\leq 2a
\end{array}
\right.
\label{pwave1d}
\end{equation}
%\begin{equation}\label{}
%\mathcal{I}(r)_{S_{ext}}=  \left\{\begin{array}{cc}
%\frac{1}{16}[6\Si(x)+4\Si(x-\alpha)+4\Si(x+\alpha)\\
%+\Si(x+2\alpha)+\Si(x-2\alpha)]-\pi/2, & r\geq 2a\cr 
%\frac{1}{16}[4\Si(x)+4\Si(x+\alpha)+\Si(x+2\alpha)\\-\Si(x-2\alpha)-4\Si(\alpha)-2\Si(2\alpha)]-\pi/2,&  r\leq 2a
%\end{array}
%\right.
%\label{swave1d}
%\end{equation}
\begin{figure*}[t!]
\vspace{0.1cm}
\centerline{
\includegraphics[width=17.5cm]{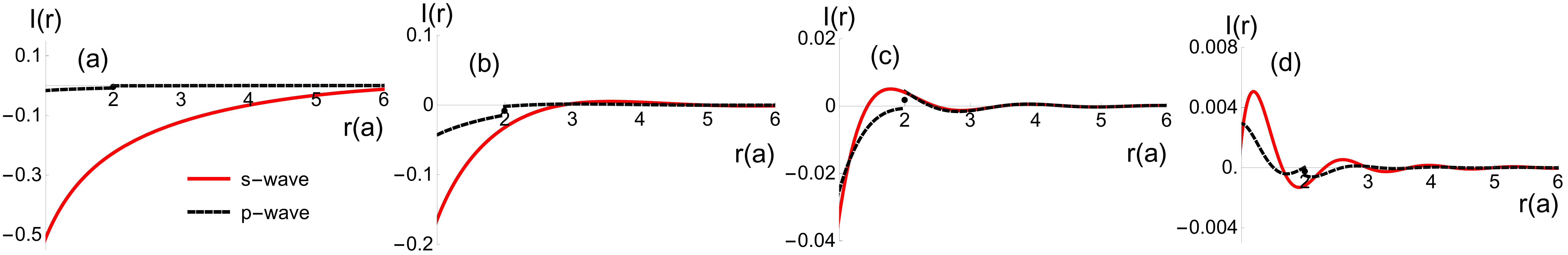}
}
\caption{
3D RKKY interactions for s-wave (red, solid), p-wave (black, dashed). $\mathcal{I}(r)$ and $r$ has units of $J_K^2(k_Fa)^6/8\pi^3 \epsilon_F$ and $a$, lattice spacing. (a) $k_Fa=\pi/10$, (b) $k_Fa=\pi/4$,(c) $k_Fa=\pi/2$ and (d) $k_Fa=3\pi/4$. The black dots correspond to $\mathcal{I}^3_p(2a)$, where there is a jump for $r>2a$ and $r<2a$ (see eq. \ref{pwave3d}).}
\label{Fig:pwave3d}
\end{figure*}

 $\mathcal{I}_p^1(r)$ has different forms for $r$ less or greater than $2a$ since some of the residues in Eq. \ref{Lindhard} change sign, resulting in a continuous but non-analytical form for the interaction. Similar effects lead to a discontinuous form in three dimensions. This non-analytical form does not lead to any unphysical properties since $\mathcal{I}(r)$ is only evaluated at discrete points.

The effective interaction between sites $i$ and $j$ in Eq. \ref{pwave1d} can be viewed as sites $(r_{i+1} - r_{i-1})$ interacting simultaneously with  sites  $(r_{j+1} - r_{j-1})$. This leads to  total of 16 processes since RKKY is a second order process. 6 of these are effectively the same site (s-wave) interaction whereas as 8 are nearest neighbor and 2 are next nearest neigbor. Now, $2k_F a$ acts as a phase shift among these and the resulting interference gives rise to rich behavior. 

For small $k_Fa \ll \pi/2$, Eq. \ref{pwave1d} gives $\mathcal{I}^1_p(r) = 0 +\mathcal{O}[(k_F r)^5]$, with the p-wave form factor at each site each providing destructive interference, almost cancelling the interaction entirely, Fig. 2(a). On the other hand, for $k_Fa=\pi/2$, the spatial phase-shift ${\rm e}^{i 2a k_F } = -1$ cancels the sign from the p-wave form factor, leading to constructive interference. Indeed s-wave and p-wave interactions are equal $\mathcal{I}^1_p(r) \simeq \mathcal{I}^1_s(r)$ for large $r$ as shown in Fig. 1(c). At small $r$ the behaviour is more complicated in this case due to lattice effects. Apart from these limiting cases, for an arbitrary $k_F$, the p-wave RKKY interaction is generally reduced compared to the s-wave case [Fig. 2(b)-(d)].

\noindent
{\it RKKY interaction in 3D -} Proceeding as above \cite{Ruderman_PRB1954},

\begin{eqnarray}
\mathcal{I}^3_s(r) = - \frac{J_K^2}{\epsilon_F}\frac{(k_Fa)^6}{8\pi^3}\frac{\sin(2k_Fr)-2k_Fr\cos(2k_Fr)}{(2k_Fr)^4}
\label{swave3d}
\end{eqnarray}

\begin{eqnarray}
\mathcal{I}_p^3(r) &=&
\begin{cases}
\frac{1}{16(k_Fr)^2}[6 f(r)-4 f(r+a)\\-4 f(r-a)+f(r+2a)+f(r-2a)], & r> 2a\\[.5cm]
\frac{1}{16(k_F 2a)^2}[4f(2a)-4f(3a)+f(4a)],& r= 2a\\[.5cm]
\frac{1}{16(k_Fr)^2}[4f(r)-4f(r+a)+f(r+2a)\\-f(r-2a)+4f(a)-2f(2a)], & r<2a
\end{cases} \nonumber \\
\label{pwave3d}
\end{eqnarray}
where $f(r) = (k_Fr)^2 \mathcal{I}_s^3(r)$. In 3D, the decay is much faster ($1/r^3$) than in 1D ($1/r$). As a result the oscillations are overdamped for small $k_F$. However, as in 1D, the p-wave case  is suppressed for small $k_F$, Fig. 3(a). On the other hand, the antiferromagnetic interaction is stabilized for large $k_F$ as shown in Fig. 3(d); again, for $k_F a = \pi/2$, $\mathcal{I}_p^3(r) \simeq \mathcal{I}_s^3(r)$ for large $r$.	

\noindent
{\it Discussion - } The RKKY interaction oscillates with $2k_Fa$ and decays with $1/r^D$. Thus, in 3D, magnetic ordering is to a first approximation determined by the nearest neighbor exchange. This is ferromagnetic for small $k_F$, turning antiferromagnetic upon increasing  $k_F$. As a result, the destructive interference for the p-wave case  is detrimental to ferromagnetism at small $k_F$, when it would otherwise be strongest. This may very well be  the reason why ferromagnetism is so rare in Kondo lattice magnets. Even though a complete list of Kondo lattice magnets is to our knowledge not available, their rareness can be gleaned by estimating there to be about 100-200 antiferromagnets\cite{Manuel_privatecomm}, including the archetypical Kondo lattice families,  Ce-115\cite{Llobet_PRB2004}, Ce-122\cite{Beyermann_PRB1991}, Yb-122\cite{Steglich_JPCM2012}. On the other hand, there are only about 10-20 Kondo lattice ferromagnets\cite{Brando_RMP2016, Manuel_privatecomm}. 
As an aside, one might expect superexchange interactions among local moments themselves to favor antiferromagnetism. However estimates for its strength yield a sub-Kelvin scale\footnote{Direct hopping among the local moments, $t_f\sim 5$ meV and Coulomb correlations, $U \sim 10$ eV lead to a mean field transition temperature scale $T_c \sim 4z t_f^2/U \sim 0.5 K.$} since the Coulomb correlations for rare earth ions are large ($U\sim 10$ eV) and their direct overlap small ($t_f \sim 1-5$ meV)--RKKY should a priori be dominant.

With the p-wave RKKY interaction doubly suppressed for small $k_F$ due to vanishing form factor at both local moments, the magnetic exchange scale is reduced, so that the Kondo effect can instead lead to heavy fermion metal formation. Indeed the suppression is more severe for higher angular momentum form factors, including d-wave, since $\Phi_k \sim k^{\Delta l}$ for small $k$. The ideal case for the enhanced ferromagnetic exchange is extended s-wave ($\Delta l = 0$) form factor\footnote{Onsite s-wave form factor is forbidden since the local moment and the conduction electron can not hybridize as they have different symmetries.}, which requires $f$ orbital conduction electrons that is not possible in real materials.

Therefore we propose to investigate materials that break inversion symmetry, at least locally at the local moment site. Since $l$ is then no longer a good quantum number, different types of form factors can mix. We argue that this is the most plausible mechanism to induce an extended s-wave form factor. Extended s-wave RKKY interactions are not reduced at small $k_F$ where ferromagnetism should be the strongest (see supplementary information for details). Indeed many ferromagnetic Kondo lattice systems have broken inversion symmetry at the local moment site including CeAgSb$_2$\cite{Sologub_JSSC1995}, CeRuPO\cite{Zimmer_JAC1995}, YbNiSn\cite{Kasaya_JPSJ1991}, YbPtGe\cite{Katoh_2008}, YbRhSb\cite{Muro_PRB2004}, YbPdSi\cite{Tsujii_JPCM2016}, YbPdGe\cite{Seropegin_JAC1995}, $\beta -$CeNiSb$_3$\cite{Thomas_IC2007}, CeTiGe$_3$\cite{Manfrinetti_SSC2005}, CeSi$_x$\cite{Hohnke_AC1966}, CePd\cite{Kappler_JLCM1985} and CePdIn$_2$\cite{Giovannini_Intmet2003}. To our knowledge, among Kondo lattice ferromagnets only Yb(Rh$_{1-x}$Co$_x$)$_2$Si$_2$\cite{Klingner_PRB2011} and YbCu$_2$Si$_2$\cite{Shimizu_JPSJ1987} have full inversion symmetry. Thus we propose screening systems with broken inversion symmetry at the local moment site as a promising route for the search of new Kondo lattice ferromagnets. 

\noindent
{\it Conclusion -} We have shown that the p-wave form factor for the Kondo interaction, common for the majority of Kondo lattice systems, is unfavorable for ferromagnetic RKKY interactions. We propose a new route for designing ferromagnetic Kondo lattice systems utilizing the broken inversion symmetry at the local moment site. More realistic calculations, including different types of crystal symmetries require future work. We believe such calculations can address other open problems in the field including magnetic ordering that is perpendicular to the easy axis\cite{Araki_PRB2003,Andrade_PRB2014} which is otherwise quite unusual but common to these materials\cite{Manuel_privatecomm}.

\bigskip
\noindent
{\it Acknowledgments - } We would like to thank Manuel Brando, Christoph Geibel and Turan Birol for fruitful discussions. This work is in part supported by the DFG via the Leibniz Prize Programme. 

%\bibliography{references}
%\bibliographystyle{apsrev}

\newpage
\onecolumngrid

\section{Supplementary information}
In the supplementary information, we provide the details of the derivation of our results.
\subsection{Canonical transformation}
We would like to derive the effective magnetic interaction among two sites, site $1$ and $2$. Therefore we consider the Kondo coupling only on these sites along with a band of conduction electrons, given as follows:
\begin{eqnarray}
H_K&=& J_K \sum_{i=1}^2 {\bf S}_i \cdot \psi_{i\alpha}^\dagger \boldsymbol{\sigma}_{\alpha \beta} \psi_{i\beta} \nonumber \\
 &=& \frac{J_K}{N}\sum_{i=1}^{2}\sum_{{\bf k},{\bf q}}e^{i(\textbf{k}-\textbf{q}) \cdot \textbf{r}_i} \textbf{S}_i \cdot  \psi_{\bf{q}\alpha}^\dagger \boldsymbol{\sigma}_{\alpha \beta} 
\psi_{\textbf{k}\beta}\\
H_0 &=& \sum_{k\alpha} (\epsilon_k-\mu) c_{k\alpha}^\dagger c_{k\alpha} 
\end{eqnarray}
where $\psi_{i\alpha}^\dagger = \Phi({\bf \hat{r}}_{ij})^{\alpha\beta} c^\dagger_{j\beta}$ as given in eq. 3 in the main text. The total Hamiltonian is the sum of the two terms, $H=H_0+H_K$. Next we carry out a canonical transformation to eliminate the $H_K$ term.
\begin{eqnarray}
\tilde{H} &=&  {\rm e}^{iS} H {\rm e}^{-iS} \nonumber \\
\tilde{H} &\simeq& (H_0+H_K) + i[S, (H_0+H_K)] \nonumber \\ &&+ \frac{i^2}{2} [S, [S, (H_0+H_K)]] + ...
\end{eqnarray}
We choose $iS$ which satisfies $[iS, H_0] = -H_K$, such that  $H_K$ term is eliminated to the lowest order:
\begin{eqnarray}
iS = \frac{J_K}{N}\sum_{i=1}^{2}\sum_{\textbf{k}, \textbf{q}} \frac{e^{i(\textbf{k}-\textbf{q}) \cdot \textbf{r}_i}}{\epsilon_{k} - \epsilon_{q}} \textbf{S}_i \cdot  \psi_{\bf{q}\alpha}^\dagger \boldsymbol{\sigma}_{\alpha \beta}\psi_{\textbf{k}\beta}
\label{iS}
\end{eqnarray}
Keeping the terms up to $J_K^2/\epsilon_K$, the effective Hamiltonian is $\tilde{H}=H_0+H_{ RKKY}$ where
\begin{eqnarray}
H_{RKKY}= \frac{1}{2}[iS, H_K] 
\end{eqnarray}
The commutator above has terms that are onsite like $S_1^{\alpha}S_1^{\beta}$ and $S_2^{\alpha}S_2^{\beta}$ which only provide a Hartree-shift as well as intersite term like $S_1^{\alpha}S_2^{\beta}$ and $S_2^{\alpha}S_1^\beta$ which mediate the magnetic exchange. Focusing on the intersite terms, different exchanges terms $S_1^{\alpha}S_2^{\beta}$ can be evaluated with through the commutator
\begin{eqnarray}
[\Psi_{q}^\dagger \sigma^{\alpha}\Psi_k, \Psi_{q^\prime}^\dagger \sigma^{\beta}\Psi_{k^\prime}]
\end{eqnarray}
where we have introduced two component vector $\Psi_k^\dagger = (\psi^\dagger_{k \uparrow }, \psi^\dagger_{k \downarrow})$ for convenience. The above commutator vanishes except $q = k^\prime $ and $k = q^\prime$. Then the non-zero terms in the commutator are
\begin{eqnarray}
H_{RKKY} =-\frac{J_K^2}{2 N^2}\sum_{i=1}^{2}\sum_{\bf kq, \alpha \beta} \frac{{\rm e}^{i ({\bf k}-{\bf q})\cdot {\bf r}_i}}{\epsilon_k-\epsilon_q}[S_i^\alpha \Psi_{q}^\dagger \sigma^{\alpha}\Psi_k, S_{\bar{i}}^\beta \Psi_{k}^\dagger \sigma^{\beta}\Psi_{q}]
\label{commutator}
 \end{eqnarray}
where $\bar{i}$ is defined as $\bar{i} \neq i$. Similarly we introduce $C_k^\dagger = (c^\dagger_{k\uparrow}, c^\dagger_{k\downarrow})$ where $\Psi_k = \Phi_k C_k$. For the p-wave form factor $\Phi_k = \sum_i^D \sin(k_i a)\sigma^i$, obtained from the Fourier transform of eq. 3 in the main text, it satisfies $\Phi_k^\dagger = \Phi_k$. Evaluating the commutator in eq. \ref{commutator} we get
\begin{eqnarray}
[S_1^\alpha C_q^\dagger \Phi_q^\dagger \sigma^\alpha \Phi_k C_k, S_2^{\beta}C_k^\dagger \Phi_k^\dagger \sigma^\beta \Phi_q C_q]&=& 
S_1^\alpha S_2^\beta \Phi_k^\dagger\Phi_k \Phi_q^\dagger\Phi_q \sigma^\alpha \sigma^\beta (C_q^\dagger C_q - C_k^\dagger C_k) \nonumber \\
&=&S_1^\alpha S_2^\beta {\rm Tr}(\Phi_k^\dagger \Phi_k){\rm Tr}(\Phi_q^\dagger \Phi_q)(i\epsilon^{\alpha \beta \gamma} \sigma^{\gamma}(n_q+n_k)+\delta^{\alpha\beta}(n_q-n_k))/4
\end{eqnarray}
where ${\rm Tr} (\Phi_k^\dagger \Phi_k)/2 = \sum_i^D \sin(ka)^2$ and $n_k=(c_{k\uparrow}^\dagger c_{k\uparrow}+ c_{k\downarrow}^\dagger c_{k\downarrow})$. For the diagonal terms $\alpha = \beta$, the contribution from $S_1^{\alpha}S_2^{\alpha}$ and $S_2^{\alpha}S_1^{\alpha}$ adds up. However for the off-diagonal terms, $\alpha \neq \beta$, the contribution from $S_1^{\alpha}S_2^{\beta}$ cancels with $S_2^{\beta}S_1^{\alpha}$ since $\epsilon^{\alpha\beta\gamma} = -\epsilon^{\beta\alpha\gamma}$. As a result, the exchange retains the symmetric Heisenberg form:
\begin{eqnarray}
H_{RKKY} = \sum_{i\neq j} \mathcal{I}(r) {\bf S}_i \cdot {\bf S}_j
\end{eqnarray}
where $\mathcal{I}(r)$ is given as
\begin{eqnarray}
\mathcal{I}(r)=-\frac{ J_K^2}{8 N^2}\sum_{\bf kq} \frac{{\rm e}^{i {\bf (k-q)\cdot r}}}{\epsilon_k-\epsilon_q}(F_q-F_k) {\rm Tr}(\Phi_k^\dagger \Phi_k){\rm Tr}(\Phi_q^\dagger \Phi_q)
\end{eqnarray} 
where we have replaced $n_k$ with the Fermi-Dirac distribution $F_k$ for finite temperatures. Next we provide the details of evaluating $\mathcal{I}$ in one and three dimensions using onsite and p-wave form factors.

\subsection{1D RKKY interaction}
\noindent
{\it S-wave - } In order to calculate $\mathcal{I}(r)$ analytically we approximate the conduction electron dispersion with a parabolic band, $\epsilon_k \simeq \hbar^2k^2/2m$. The s-wave form factor is just identity $\Phi_k = \mathbb{1}$ which gives ${\rm Tr}(\Phi_k^\dagger \Phi_k)=2$. First we evaluate the $q$ integral, 
\begin{eqnarray}
\int d\textbf{q} \frac{e^{i \textbf{q} \cdot \textbf{r}}}{q^2 - k^2} & =& \int_{0}^{\infty} dq \frac{e^{i qr}}{q^2 - k^2}  + \int_{0}^{\infty} dq \frac{e^{-i qr}}{q^2 - k^2} \nonumber\\ 
&=& \int_{-\infty}^{\infty} dq \frac{e^{i qr}}{q^2 - k^2} \nonumber \\
& =& -\frac{\pi}{k} \sin(kr)
\end{eqnarray}
Next we carry out the $k$ integral
\begin{eqnarray}
\mathcal{I}_s^1(r) &=& -\frac{J_K^2}{2} \frac{2m}{\hbar^2} \big(\frac{a}{2\pi}\big)^2  \int  d{\bf k} \frac{-\pi \sin(kr)}{k}e^{-i\textbf{k} \cdot \textbf{r}} \nonumber \\
& =& J_K^2 \frac{\pi m}{\hbar^2} \big(\frac{a}{2\pi}\big)^2  \int_0^{k_F} dk \frac{2\sin(kr)\cos(kr)}{k} \nonumber \\
& =& J_K^2 \frac{\pi m}{\hbar^2} \big(\frac{a}{2\pi}\big)^2  {\rm Si}(2k_F r) \nonumber \\
& =& \frac{J_K^2}{\epsilon_F} \frac{(a_0k_F)^2}{8\pi} {\rm Si}(2k_F r) 
\end{eqnarray}
where $a$ is the lattice spacing. The factor $(a_0k_F/2\pi)^2$ originates from converting the volume to the number of sites. ${\rm Si}$ is the sine integral function. This expression is first obtained by Ruderman and Kittel\cite{Ruderman_PRB1954} however it does not vanish at large distances $r\rightarrow \infty$. This issue arises from a strong non-analyticity at $k=0$ and $q=0$. Proper treatment\cite{Yafet_PRB1987, Litvinov_PRB1998} of this non-analyticity gives another factor of $\pi/2$ which we omit the details,
\begin{eqnarray}
\mathcal{I}_s^1(r)= \frac{J_K^2}{\epsilon_F} \frac{(a_0k_F)^2}{8\pi}[ {\rm Si}(2k_F r) - \pi/2]
\end{eqnarray}

\noindent
{\it P-wave - } The p-wave form factor in one dimensions is $\Phi_k = \sin(k_j a) \sigma_j$ where $j$ is the direction of the one dimensional chain. This results to ${\rm Tr}(\Phi_k^\dagger \Phi_k) = 2\sin(k)^2$, where we dropped the $j$ index since the results are independent of it. Then we need to evaluate 
\begin{eqnarray}
\mathcal{I}_p^1(r)=-\frac{ J_K^2}{2N^2}\sum_{\bf kq} \frac{{\rm e}^{i {\bf (k-q)\cdot r}}}{\epsilon_k-\epsilon_q}\sin(k)^2\sin(q)^2(F_q-F_k) 
\end{eqnarray} 
Similar to s-wave case, we first carry out the $q$ integral,
\begin{eqnarray}
\int d\textbf{q} \frac{\sin(qa)^2 e^{i \textbf{q} \cdot \textbf{r}}}{q^2 - k^2} & =& \int_{0}^{\infty} dq \frac{\sin(qa)^2 e^{i qr}}{q^2 - k^2}  + 
\int_{0}^{\infty} dq \frac{\sin(qa)^2 e^{-i qr}}{q^2 - k^2} \nonumber\\ 
&=& \int_{-\infty}^{\infty} dq \frac{\sin(qa)^2 e^{i qr}}{q^2 - k^2} \nonumber \\
&=&
\begin{cases}
\frac{\pi}{4k}[-2\sin(kr)+\sin(k(r+2a))+\sin(k(r-2a))],& ~~{\rm for ~}r\geq 2a\\[0.2cm]
\frac{\pi}{4k}[-2\sin(kr)+\sin(k(r+2a))-\sin(k(r-2a))], &~~{\rm for ~}r\leq 2a
\end{cases}
\end{eqnarray}
where the residues changes sign for $r$ less or greater than $2a$ for the $\sin(k(r-2a))$ term. Next we carry out the $k$ integral,
\begin{eqnarray}
\mathcal{I}_p^1(r) &=& -\frac{J_K^2}{2}\frac{2m}{\hbar^2} \big(\frac{a}{2\pi}\big)^2  \int  d{\bf k} \frac{\pi (-2\sin(kr)+\sin(k(r+2a))\pm \sin(k(r-2a)))}{4k}\sin(ka)^2e^{-i\textbf{k} \cdot \textbf{r}} \nonumber \\
&=& \frac{J_K^2}{8}\frac{ 2m\pi}{\hbar^2} \big(\frac{a}{2\pi}\big)^2  \int_0^{k_F}  dk \frac{(2\sin(kr)-\sin(k(r+2a))\mp \sin(k(r-2a)))}{k}2\sin(ka)^2\cos(kr)\nonumber \\
&=&
\begin{cases}
\frac{1}{16}\frac{J_K^2}{\epsilon_F}\frac{(ak_F)^2}{8\pi}[ 6 {\rm Si}(2 k_F r)-4 {\rm Si}(2 k_F (r-a)) - 4 {\rm Si}(2 k_F (r+a)) \nonumber\\
+{\rm Si}(2 k_F (r+2a))+{\rm Si}(2 k_F (r-2a))],& ~~{\rm for}~r\geq2a \\
\frac{1}{16}\frac{J_K^2}{\epsilon_F}\frac{(ak_F)^2}{8\pi}[ 4 {\rm Si}(2 k_F r) - 4 {\rm Si}(2 k_F (r+a))+{\rm Si}(2 k_F (r+2a)) \nonumber \\-{\rm Si}(2 k_F (r-2a))
+2{\rm Si}(4k_Fa)-4{\rm Si}(2k_Fa)],& ~~{\rm for}~r\leq2a
\end{cases}
\end{eqnarray}
Note that there is no extra factor of $\pi/2$ since the non-analyticity at $k = 0$ and $q= 0$ is regulated by the $\sin(k)^2\sin(q)^2$ term. The above expression can be expressed in terms of the s-wave interaction, $\mathcal{I}^1_s$
\begin{equation}\label{}
\mathcal{I}_p^1(r)=  \left\{\begin{array}{cc}
\frac{1}{16}[6\mathcal{I}_s^1(r)-4\mathcal{I}_s^1(r-a)-4\mathcal{I}_s^1(r+a)
+\mathcal{I}_s^1(r+2a)+\mathcal{I}_s^1(r-2a)], & r\geq 2a\\[0.15cm] 
 \frac{1}{16}[4\mathcal{I}_s^1(r)-4\mathcal{I}_s^1(r+a)+\mathcal{I}_s^1(r+2a)+\mathcal{I}_s^1(2a-r)-4\mathcal{I}_s^1(a)+2\mathcal{I}_s^1(2a)],& r\leq 2a
\end{array}
\right.
\label{pwave1d_app}
\end{equation}
\subsection{3D RKKY interaction}
{\it S-wave -} Similar to 1D, we evaluate the $q$ integral first
\begin{equation}
\int d\textbf{q} \frac{e^{i \textbf{q} \cdot \textbf{r}}}{q^2 - k^2} = 2\pi \int_{0}^{\infty}q^2 dq \int_{0}^{\pi} d\theta \sin\theta\frac{\rm{e}^{iqr \cos\theta}}{q^2 - k^2} = \frac{2\pi^2}{r} \cos(kr)
\end{equation}
where $\theta$ is the angle between ${\bf r}$ and ${\bf q}$. Next we carry out the $k$ integral
\begin{eqnarray}
\mathcal{I}_s^3(r) &=& \frac{J_K^2}{2}\frac{2m}{\hbar^2}\big(\frac{a}{2\pi}\big)^6\int d {\bf k }\frac{2\pi^2}{r} \cos(kr) {\rm e}^{i {\bf k}\cdot {\bf r}} \nonumber \\
&=& -\frac{J_K^2}{2}\frac{2m}{\hbar^2}\big(\frac{a}{2\pi}\big)^6 \frac{4\pi^3}{r} \int_0^{k_F} dk~ k^2 cos(kr) \int_0^\pi d\theta \sin(\theta) {\rm e}^{i kr \cos\theta} \nonumber \\
&=& -\frac{J_K^2}{2}\frac{2m}{\hbar^2}\big(\frac{a}{2\pi}\big)^6 \frac{8\pi^3}{r^2} \int_0^{k_F} dk ~ k \sin(kr) \cos(kr) \nonumber \\
&=& -\frac{J_K^2}{\epsilon_F}\frac{(k_Fa)^6}{8\pi^3}\frac{\sin(2k_Fr)-2k_Fr\cos(2k_Fr)}{(2k_Fr)^4}
\end{eqnarray}

\noindent
{\it P-wave -} P-wave form factor in three dimensions takes the form, $\Phi_k = \sum_{i=x,y,z} \sin(k_i)\sigma^i$. Then the trace of the form factors simplify to ${\rm Tr}(\Phi_k^\dagger \Phi_k) = 2\sum_{i=x,y,z} \sin(k_i)^2 \sim 2\sin(k)^2$. Carrying out the $q$ integral, 

\begin{eqnarray}
\sum_{q}\frac{e^{-i\textbf{q} \cdot \textbf{R}}}{q^2-k^2}\sin^2(qa) &=& 2\pi \int_0^{\infty}dq~ \frac{q^2 \sin(qa)^2}{q^2-k^2} \int_0^\pi d\theta {\rm e}^{i qr \cos \theta} \nonumber \\
&=&
\begin{cases}
\frac{\pi^2}{2r}\Big[2\cos(kr)-\cos(k(r+2a)) - \cos(k(r-2a))  \Big], & r > 2a\\[0.2cm] 
\frac{\pi^2}{2r}\Big[2\cos(kr)-\cos(k(r+2a))  \Big], & r = 2a\\[0.2cm] 
\frac{\pi^2}{2r}\Big[2\cos(kr)-\cos(k(r+2a)) + \cos(k(r-2a)) \Big], & r > 2a
\end{cases}
\end{eqnarray}
Next we carry out the $k$ integral,
\begin{eqnarray}
\mathcal{I}_p^3(r) &=& -\frac{J_K^2}{2}\frac{2m}{\hbar^2}\big(\frac{k_Fa}{2\pi}\big)^6\frac{\pi^2}{2r} \int d{\bf k} \sin(ka)^2 \big(2\cos(kr)-\cos (k(r+2a))\mp \cos(r-2a)\big){\rm e}^{i{\bf k}\cdot {\bf r}} \nonumber\\
&=&-\frac{J_K^2}{2}\frac{2m}{\hbar^2}\big(\frac{k_Fa}{2\pi}\big)^6\frac{2\pi^3}{r^2} \int_0^{k_F} dk~k \sin(ka)^2 \sin(kr)\big(2\cos(kr)-\cos (k(r+2a))\mp \cos(r-2a)\big) \nonumber \\
&=&
\begin{cases}
\frac{1}{16(k_Fr)^2}[6 f(r)-4 f(r+a)-4 f(r-a)+f(r+2a)+f(r-2a)], & r> 2a\\[.15cm]
\frac{1}{16(k_F 2a)^2}[4f(2a)-4f(3a)+f(4a)],& r= 2a\\[.15cm]
\frac{1}{16(k_Fr)^2}[4f(r)-4f(r+a)+f(r+2a)-f(r-2a)+4f(a)-2f(2a)], & r<2a
\end{cases}
\end{eqnarray}
where $f(r) = (k_Fr)^2\mathcal{I}_s^3$.
\subsection{Extended s-wave form factor}
The extended s-wave form factor in one dimension where the form factor couples nearest neighbors with the same sign, $\psi_{i \alpha}^\dagger = (-i/2)(c_{i+1\alpha}^\dagger+c_{i-1\alpha}^\dagger)$, as shown in Fig. 1(b) in the main text, leads to ${\rm Tr}(\Phi_k^\dagger \Phi_k) = 2 \cos(ka)^2$. Carrying out similar calculations, we get
\begin{eqnarray}
\mathcal{I}_{s_{ext}}^1(r) &=&\begin{cases}
\frac{1}{16}\frac{J_K^2}{\epsilon_F}\frac{(ak_F)^2}{8\pi}[ 6 {\rm Si}(2 k_F r)+4 {\rm Si}(2 k_F (r-a)) + 4 {\rm Si}(2 k_F (r+a)) \nonumber\\
+{\rm Si}(2 k_F (r+2a))+{\rm Si}(2 k_F (r-2a))-8\pi],& ~~{\rm for}~r\geq2a \\
\frac{1}{16}\frac{J_K^2}{\epsilon_F}\frac{(ak_F)^2}{8\pi}[ 4 {\rm Si}(2 k_F r) + 4 {\rm Si}(2 k_F (r+a))+{\rm Si}(2 k_F (r+2a)) \nonumber \\+{\rm Si}(2 k_F (r-2a))
+2{\rm Si}(4k_Fa)+4{\rm Si}(2k_Fa)-8\pi],& ~~{\rm for}~r\leq2a
\end{cases}
\end{eqnarray}
where the main difference compared to the p-wave form factor is that all terms contribute with a plus sign. As a result there is no destructive interference for small $k_F$ and ferromagnetic RKKY interaction is large (Fig 4).

\begin{figure}[t!]
\vspace{0.1cm}
\centerline{
\includegraphics[width=12cm]{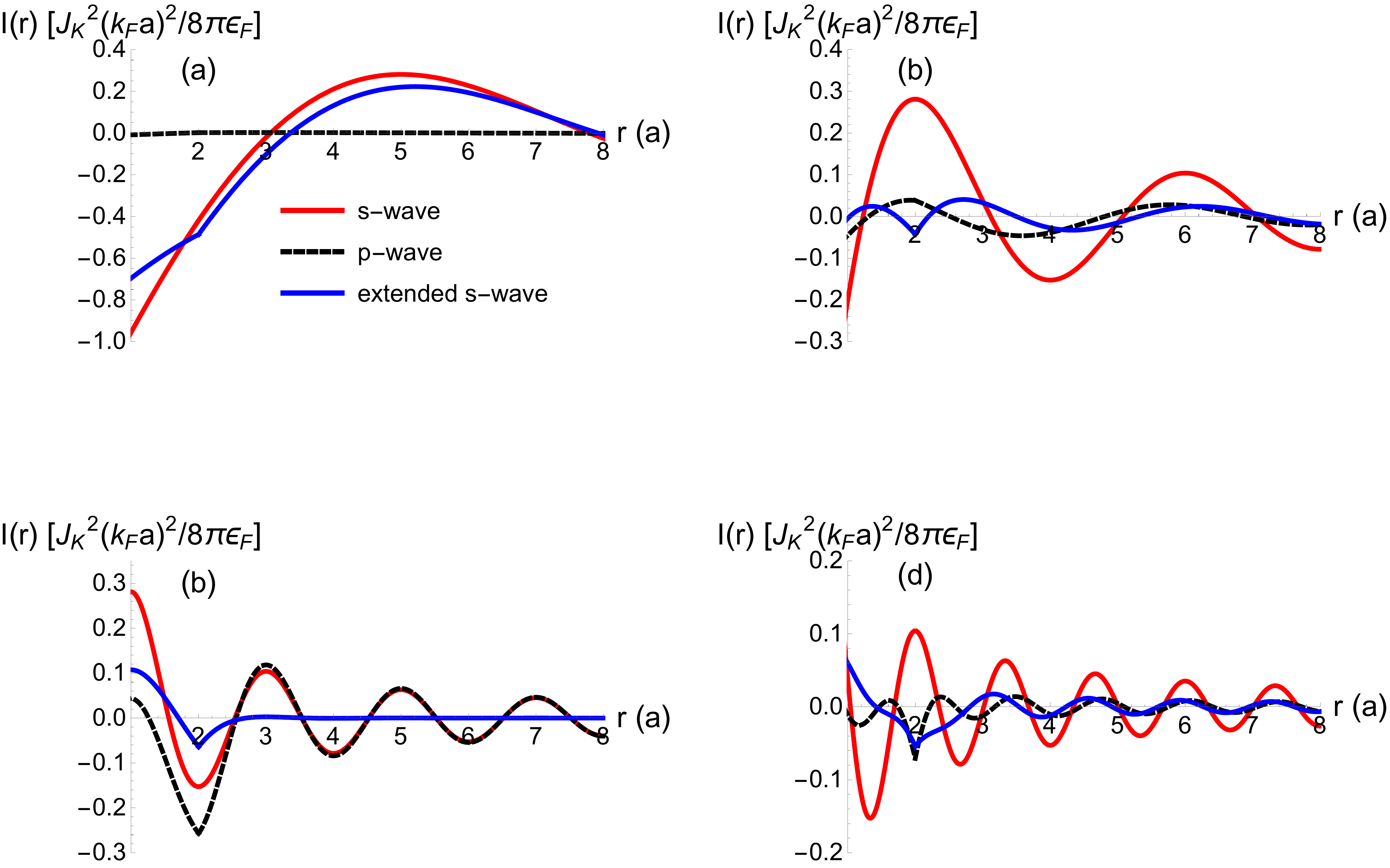}
}
\caption{1D RKKY interactions for s-wave (red, solid), p-wave (black, dashed) and extended s-wave (blue, solid) for (a) $k_F=\pi/10$, (b) $k_F=\pi/4$,(c) $k_F=\pi/2$ and (d) $k_F=3\pi/4$. There is no destructive interference for extended s-wave case at small $k_F$. As a result, the ferromagnetic RKKY interaction is large.}
\label{sext}
\end{figure}

%\bibliography{references}
%\bibliographystyle{apsrev}

\end{document}